# Personalization of learning through adaptive technologies in the context of sustainable development of teachers' education

*Maiia* Marienko[1,*], *Yulia* Nosenko[1], *Alisa* Sukhikh[1], *Viktor* Tataurov[2], and *Mariya* Shyshkina[1]

[1]Institute of Information Technologies and Learning Tools of the NAES of Ukraine, Department of Cloud-Oriented Systems of Education Informatization, Kyiv, Ukraine
[2]Kamianets-Podilskyi National University named after Ivan Ogienko, Department of Computer Sciences, Kamianets-Podilskyi, Ukraine

**Abstract.** The article highlights the issues of personalized learning as the global trend of the modern ICT-based educational systems development. The notion, the main stages of evolution, the main features and principles of adaptive learning systems application for teachers' training are outlined. It is emphasized that the use and elaboration of the adaptive cloud-based learning systems are essential to provide sustainable development of teachers' education. The current trends and peculiarities of the cloud-based adaptive learning systems development and approach of their implementation for teachers' training are considered. The general model of the adaptive cloud-based learning system structure is proposed. The main components of the model are described; the issues of tools and services selection are outlined. The methods of the cloud-based learning components introduction within the adaptive systems of teacher training are considered. The current research developments of modeling and implementation of the adaptive cloud-based systems are outlined.

## 1 Introduction

In the context of modern transformations of social and technological systems, ubiquitous digitization, rapid updating of content and contexts of learning, the emergence of new scientific facts, changes in professional standards, requirements and expectations of employers, a social demand is formed for educators who can constantly improve their competence for sustainable personal development. Training a new generation of educators able to work in a dynamic environment, adapt to constant change, form subject-subject relationships in the classroom, including the use of modern ICT, is an important task of educational institutions.

As stated in the Digital Agenda for Europe [1], about 90% of occupations currently require at least basic ICT skills. The complexity of the tasks assigned to teachers is since they must not only develop their digital competence, but also the competence of the students. They also need to develop the skills for self-development throughout life.

The modern demands for educators imply they should be able to select appropriate and apply effectively the emerging technologies in the educational process that allow personalizing the educational process as much as possible, to bring it closer to the learning needs of each student: multimedia resources, electronic educational game resources, adaptive technologies, etc. The constant updating of these technologies requires the modern teacher to be reflexive, able to critically evaluate one's abilities, to direct efforts for self-development and self-improvement.

Teachers' training is fundamentally taken in the context of lifelong learning. According to Organization for Economic Cooperation and Development (OECD) experts, about 94% of teachers from participating countries are involved in at least one professional development activity during a year. Often, such training takes place online, through non-formal or informal learning, during off-hours. In this regard, there is a need to integrate technologies that will maximize the effectiveness of teachers' competences development while minimizing time spent. Progressive in this sense are adaptive technologies that allow for personalization of learning and significantly improve the quality of education, optimize time and other resources.

The important accelerator of the progress in this field is the availability of the cloud-based platforms and tools that force the emergence of a new generation of adaptive learning systems.

The cloud-based adaptive networks and platforms provide the united framework for the integration of different kinds of educational services into the whole system. Just the cloud-based infrastructures and services that are aimed to provide the openness and flexibility of systems' design are the most appropriate for the sustainable development of teachers' education. The features of the cloud-based systems are to support the flexible selection and adaptive adjustment of their components for certain groups of learners' needs. These systems may be implemented for different subject domains, using the uniform platform and also, they are open for innovations and the current introduction of new and emerging technologies.

---

* Corresponding author: popelmaya@gmail.com

The cloud-based platforms are invariant as for the content of methodical systems realized by these platforms so the basic learning and research components may be revealed for the possible general prototype of such systems. The important issue is to design the system to fit most of all the aims of teachers' sustainable education and professional development. This caused the need to consider the general structure of the adaptive cloud-based system of teachers' education and reveal its basic components having in mind that the selection of services for each of the components and elaboration of methods of their use may be the separate task for the appropriate context.

*The purpose of the article* is to consider the state of the art, tools, and services of the cloud-based adaptive learning systems introduction in the context of sustainable development of teachers' training, outline the approach, the general model, the methods and prospective trends of their use.

## 2 Personalization of learning as a global educational trend

The formation of a cloud-based learning and research environment is a promising area of informatization of education, which is recognized as a priority by international educational community [2, 3, 4], and is being intensively developed in various fields of education, in particular, in the teaching of mathematics and informatics disciplines in educational institutions. Trends of the introduction of cloud technologies in the educational process for organizing access to software used for various types of teamwork, in carrying out scientific and educational activities, research and development, implementation of projects, exchange of experience, etc. have become particularly relevant [3, 5].

According to many Ukrainian scientists, the problem of modernization of the teaching process in secondary school is in line with the modern scientific and technological progress [6, **Error! Reference source not found.**]. In this concern, one of the main factors for improving the quality of training of pedagogical, scientific and pedagogical staff, the widespread use of innovative pedagogical technologies is the introduction of adaptive cloud-oriented systems in educational institutions [6, **Error! Reference source not found.**].

Personalization of learning is currently one of the world's leading educational trends [8, 9, **Error! Reference source not found.**, 11, 12, 13] and more.

Personalized learning is a pedagogical concept supposing that the emphasis in the educational process is shifted from the normative requirements, standards, etc. to the personality of a learner, taking into account his / her characteristics: inclinations, abilities, and talents, national and cultural context, etc.

Having analyzed the materials of the International Academy of Education, UNESCO [14], Microsoft (Microsoft Education Transformation Framework) [15],

we highlight the main principles of personalized learning:

- *Active involvement*. It is the transition from the passive perception of information by students to problematic search, independence, and creativity.

- *Social participation*. It means the involvement of students in active participation in various types of group work, problematic discussions, group project activities, etc.

- *Meaningful activities*. This is linking the content of education to real life, drawing practice-oriented analogies, performing tasks related to solving everyday problems, etc.

- *Relating new information to prior knowledge.* It means that it is important to relate new information to previously acquired so that it does not displace acquired knowledge, but expand, deepen, facilitate the establishment of logical, interdisciplinary links.

- *Being strategic*. In today's world of super-fast updating and dissemination of information, it is important to develop the younger generation's self-study skills, which lie not only in the learning itself but also in the ability to plan and control this process, build effective strategies, own educational trajectories of life, find educational information and critically evaluate it, establish cause and effect relationships, etc.

- *Engaging in self-regulation and being reflective*. It means forming the need and knowledge to carry out self-evaluation, critically assess one's own level of knowledge, competence, using various means (including adaptive assessment), monitor learning outcomes, plan activities to fill in the identified gaps, etc.

- *Restructuring prior knowledge*. It is in avoiding that prior knowledge and experience interfere with the acquisition of new educational material (for example, to explain to children that the Earth is "round", whereas the human eye perceives it differently). It is advisable to introduce new knowledge based on already known facts and to support them with scientific evidence.

- *Understanding rather than memorization.* It is important to orient students to understand the general principles, concepts, instead of mechanically memorizing individual facts and algorithms. This is a condition for effective learning. Leading international monitoring studies (such as PISA) are built on this principle – not just the knowledge of the curriculum, but their understanding, ability to apply the knowledge effectively in practice.

- *Time for practice*. It is important to take the time to practice in the learning process to consolidate one's skills and improve them. Forming individual skills (such as reading and writing) is impossible at all without long-term management.

- *Developmental and individual differences*. In the educational process it is necessary to account for the individual characteristics of each student. These features can both enhance the student's ability (e.g., aptitude for the sciences, disjointed spatial thinking, etc.), and may impose some limitations (for example, for a student with special needs). Therefore, it is important to offer a wide range of tasks and formats, including self-assessment and self-evaluation that will motivate each student. This principle is best met by adaptive technologies.

- *Creating motivated learners*. Students' interest, their motivation in getting high results is a prerequisite for successful learning. Therefore, it is important to help in the proper arrangement of accents, goals, to choose tasks of adequate complexity (not too simple and not too complicated), to create situations of success, to avoid comparisons of students with each other, to support, including emotional factors, etc.

As we can see, the principles recognized by world experts (in particular, UNESCO) are more in line with the classical didactic principles, on the one hand, and reflect the genesis of approaches to learning: from passive to active participation, from routine memorization to creative knowledge, from the role of the teacher-translator of knowledge to the teacher-facilitator, from the precise use of technologies to their organic integration into the educational process.

In the context of the personification of learning, the educational process is designed following the principles outlined with the use of technologies that facilitate their effective implementation. Generally, truly personalized learning appears to be possible regarding the rapid development of the IT field. The development of technology has naturally led to the development of education on the way of personification, acting as a kind of catalyst and an integral part of this process. Means of support for the personification of the educational process include adaptive technologies that adjust the tools, resources and the content to the student needs in real-time, as well as provide him and the teacher with the analytics of the educational process.

# 3 Evolution of technologies for personalized learning

An important feature of adaptive cloud-oriented systems is the ability to dynamically supply computing resources and software and hardware, and its flexible customization to the needs of the user. With this approach, access to various types of educational software is organized, which can be either installed specifically on a cloud server or provided as a public service (available on any other electronic data media available via the Internet) [3]. Therefore, it is necessary to study the question: what are the ways and models of pedagogical activity, how are the role of electronic educational resources (EER) and approaches to their design changing, what are the means, models and ways of organizing access to them given the existing trends of cloud-oriented environment design and use in educational institutions. In this regard, the definition of conceptual backgrounds, features of creation and application, trends and ways of implementation of this class of systems requires careful analysis.

The adaptability of educational systems is achieved through the use of technologies that enable the automatic adaptation of these systems to the educational needs of different categories of users or the individual characteristics of learners. They can be customized depending on the level of education; educational role (student, teacher, researcher, etc.); the level of educational achievements; personal abilities, giftedness; educational needs (including special needs), etc.

To implement the computer-processing functions of a cloud-oriented system (content-technology and information-communication), a virtualized computer-technology (corporate or hybrid) infrastructure must be purposefully created.

Personalization is ensured by the possibility of customizing the ICT infrastructure (including the virtual one) to the individual information-communication, information-resource and operational-processing needs of the participants of the educational process.

Knowledge modeling tools and approaches developed in the field of artificial intelligence (AI) provide new applications in the design of computer-based training systems in connection with the development of such promising technologies as distributed knowledge bases; data repositories and shared knowledge; multi-agent technologies that enable the collective solution of tasks in a multi-user environment that communicate with each other as they exchange information and interact with software agents to support many intelligent functions.

The main stages of the evolution of adaptive cloud-oriented systems in education are shown in Table 1.

**Table 1.** The main stages of the development of adaptive learning systems.

| Stage name | Period | Computer systems implementations | The role of modeling in stage formation |
|---|---|---|---|
| Programmed learning | 1960s | Low-level programming languages (assembly language) | Thinking models in the form of algorithms |
| Educational programs | 1960s - early 1970s | High-level programming languages (BASIC, Pascal, Algol, C), GUI | Black box thinking models |
| Educational systems of artificial intelligence | the late 1970s | Artificial intelligence languages (Prolog, Lisp, etc.) | Thinking models based on knowledge representation |
| Simulation of knowledge modeling, adaptive control | 1980s – 2010s | Artificial intelligence Languages, object-oriented programming languages (C++, Visual Basic, etc.), multimedia | Simulation models of thinking and knowledge |
| Adaptive cloud-based systems | 2010 is our time | Server virtualization hardware; adaptive networks (linguistic (Semantic Web), intelligent network agents, robots, etc.) | Combining knowledge models and their representation in adaptive networks |

With the use of cloud technologies, the amount of computing power is increasing significantly, and the information and analytical tools that can be used to collect and process data that characterize student activity are being improved. The emergence in the last decades of methods of programming natural language dialogue, strategic planning, and teacher modeling testifies to the emergence of a separate phase, which is defined as ATM (Adding a tutorial model), computer systems with the teacher model [16].

It can be assumed that the further development of computer-aided learning will be in favor to improve the knowledge models that underpin it [17]. That is, as these tools acquire an increasing degree of intellectualization, they will increasingly approach the modeling of more or less holistic fragments of the educational space and particular types of educational interaction.

# 4 Current trends in the development and use of adaptive learning systems in teacher education

Adaptive software and platforms with several benefits have been tested in various educational and socio-cultural settings and are now widely used in the global educational space:

- *curriculum platforms:* Alta[a], Cerego[b], Fishtree[c], Fulcrum Labs[d], LearnSmart[e], RedBird Advanced Learning[f], Socrative[g], Smart Sparrow[h]);
- *adaptive learning management systems (LMS), creation of training courses* (Neo LMS[i], Open Learning Initiative (OLI)[j]);
- *adaptive testing systems* (Typeform[k], Quizalize[l]);
- *adaptive math's training programs for elementary school students* (DreamBox[m], i-Ready[n], LearnBop[o], SuccessMaker[p], ScoatPad[q], Splash Math[r]);
- *adaptive mathematics training programs* (KnowRe[s], LearnBop[t], Matific[u], ST Math[v], Think Through Math[w]);

- *Adaptive Adult Learning Platforms* (Elevate).

Today, Knewton is one of the most recognized and effective adaptive learning programs in the world. It is based on a nonlinear knowledge graph that connects concepts; analytics for teachers and students; personalized recommendations for teachers and students. In addition to the results of the assignments, the system also allows you to determine student's proficiency ("skill"), engagement ("involvement"), active time (how much student spent time to complete the task), time to complete the material to the end. The system generates statistics for both student (personal progress) and educator (individual student and group progress). Besides, based on the data obtained, the system provides recommendations: on which topics to work more carefully (for the student), what tasks for which topics it is advisable to offer for better mastering the material (for the educator) [18]. Arizona State University, which has been using Knewton technology since 2011, noted that the dropout rate dropped from 13% to 6% and those who completed the study increased from 66% to 75% [19], which attests to their effectiveness.

In the United States, Synaptic Global Learning, in collaboration with the Center for Innovation and Excellence in eLearning at the University of Massachusetts (USA), developed the world's first adaptive MOOC (Comprehensive Molecular Online Course) in Computational Molecular Dynamics, called aMOOC. The aMOOC platform provides powerful pedagogical support and a personalized learning environment based on Amazon Web Services cloud architecture [20].

In the US at Aspire Public Schools, teachers use adaptive technologies as a means of supporting blended learning. They use digital tools, discuss, learn about their achievements, participate in a strategy to improve their outcomes and overcome challenges [21].

Studying this problem, it is necessary to characterize to what (or to whom) these systems should be adapted, what characteristics should be investigated and taken into account when constructing a user model. In addition to the user model, the system also stores the user profile. The user profile stores personal information of users such as scientific (educational) benefits, training mode, and user knowledge. The model is based on profile research. A group of scientists from Croatia [16] explored the characteristics issues required when building a user model for adaptive learning systems. According to the research, individual users' characteristics were chosen as sources of adaptation. The result is a list of 17 characteristics that are considered sources of adaptation (age, gender, cognitive abilities, such as processing speed, long-term memory, spatial abilities, etc., metacognitive abilities, personality, anxiety, emotional and affective states, cognitive styles, learning styles, experience, background knowledge, motivation, expectations). According to the results, the adaptation of training systems increases when they are adapted to one or more of the following user characteristics.

[a] Alta: https://www.knewton.com/the-power-of-altas-adaptive-technology/
[b] Cerego: https://www.cerego.com/
[c] Fishtree: https://www.fishtree.com/
[d] Fulcrum Labs: https://www.fulcrumlabs.ai/
[e] LearnSmart: https://services.learnsmartsystems.com/sso/
[f] RedBird Advanced Learning:
https://www.edsurge.com/product-reviews/redbird-advanced-learning-courses
[g] Socrative: https://socrative.com/
[h] Smart Sparrow: https://www.smartsparrow.com/what-is-adaptive-learning/
[i] Neo LMS: https://www.softwareadvice.com/lms/neo-lms-profile/
[j] Open Learning Initiative (OLI): https://oli.cmu.edu/
[k] Typeform: https://www.typeform.com/product/
[l] Quizalize: https://www.quizalize.com/
[m] DreamBox: https://www.dreambox.com/adaptive-learning
[n] i-Ready: http://i-readycentral.com/articles/how-does-the-i-ready-adaptive-diagnostic-work/
[o] LearnBop: https://www.learnbop.com/
[p] SuccessMaker:
https://mypearsontraining.com/products/successmaker
[q] ScoatPad: https://www.scootpad.com/
[r] Splash Math: *www.splashmath.com*
[s] KnowRe: *www.knowre.com*
[t] LearnBop: *www.learnbop.com*
[u] Matific: *www.matific.com*

[v] ST Math: *www.stmath.com*
[w] Think Through Math: *content.thinkthroughmath.com*

According to English scholars [22] adaptability is a way of constructing a system of courses to model the interests of the user and apply it to adaptation based on the user's preferences. An adaptive learning system is a learning system that adapts the structure of the learning content to the individual learning characteristics of individual users.

In this regard, several important trends can be identified that characterize the promising avenues for the development and use of artificial intelligence and knowledge-based approaches in teacher education in the future:

- "intellectualization" of all units of educational systems, their further integration in the educational process and learning environment;

- intensive development and implementation of educational systems based on the latest achievements, methods, and developments of the AI industry;

- further unification, universalization, the formation of common standards for the development and implementation of individual modules, subsystems, and systems of educational purpose within the qualitatively new information and educational space with elements of artificial intelligence;

- increasing the role of the Big Data approach for collecting and analyzing the results of tracking learning processes and the individual progress of the learner;

- the increasing saturation of the learning environment with a variety of intelligent devices, remote controls, robots, peripheral equipment and the like, which can be managed on a single platform, over the network ("Internet of Things");

- increasing the role of computer literacy and technological culture of all participants in the learning process for the successful development and implementation of new generation AI learning tools.

## 5 Current research developments and future prospects

From 2018 at the Institute of Information Technologies and Learning Tools of the National Academy of Educational Sciences of Ukraine (Ukraine), a planned scientific study "Adaptive cloud-based system of secondary schools' teachers training and professional development" (2018-2020) is held.

The urgency of the work is due to the need to modernize the teaching process in the secondary school, bringing it into line with modern scientific and technological advances, which is the key to training highly qualified, ICT-competent teachers.

The purpose of the work is theoretical substantiation and development of an adaptive cloud-oriented system of education and professional development of secondary school teachers.

The main research results obtained during the implementation of the research, cover the following provisions:

1. The general model of an adaptive cloud-based system of education and professional development of teachers is proposed (Fig. 1). The model contains the components of corporate cloud of educational institution (databases and repositories, data analytics and pattern recognition tools, specialized educational software and services; educational robots, language processing tools, and others); public cloud services (usually it may contain office services; adaptive data and knowledge management tools, special learning and research cloud-based tools, electronic educational resources (EER) collections and EER elaboration tools and others; communication services (that may be either public or corporate); as well as the services of publicly available scientific-educational information networks and infrastructures. The separation of services between public and corporate cloud is conditional as it depends on services availability and educational needs.

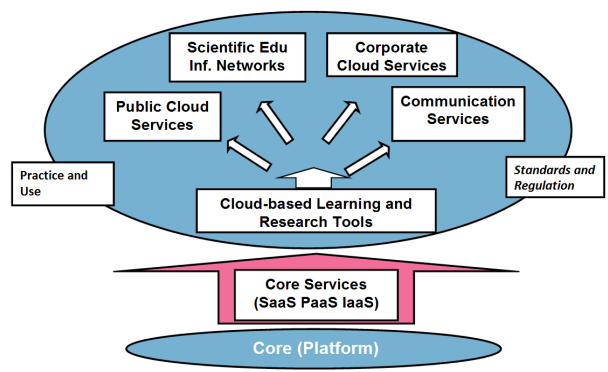

**Fig. 1.** The general structure of the adaptive cloud-based system of teachers' education and professional development.

The formation of modern cloud-based systems for supporting teaching and research activities and students' collaboration on the learning process should be based on appropriate innovative models and methodology that can ensure a harmonious combination and embedding of various networking tools into the information-educational environment of higher education. The model of the CoCalc use to support various forms of collaboration in the process of training of pre-service teachers of mathematics (Fig. 2) consists of the following components: the organizational, the content and the technological ones.

The model was built on the basis of the general structure of the adaptive cloud-based system of teachers' education and professional development. The model includes teachers and students individual or collective spaces consisting of some structural elements. The combination of certain elements provides different types of interaction organization (between students and a teacher and between students with each other) among them such as individual and group work of students, students and teachers cooperation and active communication.

2. The methods for using the services of an adaptive cloud-based system of education and professional development of teachers, including the methods of using the services of a research-educational cloud based on Microsoft Office 365 used for searching, submitting and processing data and information in open systems of study and research [5]; the method of using adaptive content management systems based on public cloud (Google Docs, IBM Box, Microsoft Office 365) to support collaboration in virtual teams; the methodology for supporting the processes of adaptive creation and use of electronic educational resources (WPadV4, AWS) are proposed. The last method is based on the utility model of using knowledge-management tools to design electronic educational resources proposed by S. Svetsky, O. Moravcik [23].

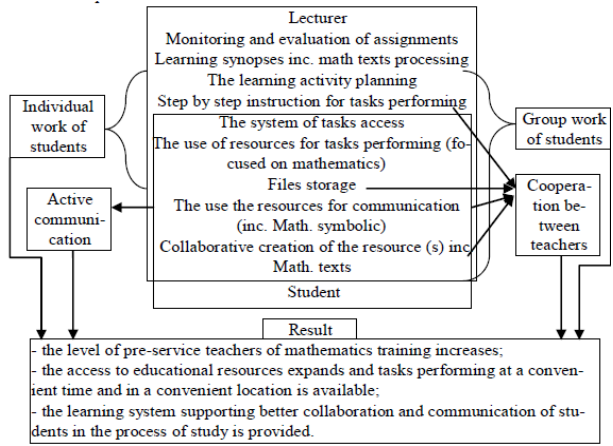

**Fig.2.** The model of the CoCalc use to support various forms of collaboration in the process of training of pre-service mathematics teachers.

To approve the application of different components of the adaptive learning system use the pedagogical experiment was undertaken. The experimental base of research was Kamyanets-Podilskyi National University named after Ivan Ogienko (Ukraine), masters of physical, mathematical, pedagogical specialties of a pedagogical institution of higher education, sample size – 160 students. The method of using Microsoft Office 365 tools, in particular, Microsoft Teams services to support students' collaboration and communication in the learning process and project elaboration was introduced. The Microsoft Office 365 was selected due to the complex of different educational oriented tools available in this package. It was essential to provide adaptability of the learning process and introduction of personalized learning techniques, in particular, while managing the students' team work. An alternative to using individual Microsoft Office 365 services may be FaceTime, for Apple gadgets - Google Duo, Hangouts. It was substantiated that due to the scientifically grounded and methodologically sound introduction of this method, there was a statistically significant increase in the level of ICT-competence and student performance (statistically consistent) [5].

3. It is substantiated that when designing adaptive cloud-based systems in a pedagogical university, it is advisable to use computer-based adaptive systems and platforms that have several advantages, tested in different educational and socio-cultural environments and are now widely used in the world educational space:

curriculum platforms (Alta, Cerego, Fishtree, Fulcrum Labs, LearnSmart, RedBird Advanced Learning, Smart Sparrow, Socrative); adaptive learning management systems, creation of training courses (Neo LMS, Open Learning Initiative (OLI)); adaptive testing systems (Typeform, Quizalize); adaptive adult learning platforms (Elevate) and more.

4. It is determined that it is expedient to include cloud services of open science, in particular, services of European research infrastructures, in the composition of facilities and services of forming adaptive cloud-oriented systems in a pedagogical university; scientific and educational networks; cloud data collection, submission and processing services; as well as the services of the European Open Science Cloud.

5. On the basis of the proposed research the model of using CoCalc cloud service as a tool of forming professional competencies of the mathematics teachers was approved, taking into account the links between the components of professional competencies and all cycles of the disciplines of the mathematics teachers' training programs for pedagogical universities. The use of the cloud service CoCalc proved to be effective at three stages of the development of professional competencies. It has been found out that the use of this cloud service in the process of pre-service training of mathematics teachers affects first of all the formation of special professional competencies.

The analysis of the results of the forming stage of the pedagogical experiment showed that the distribution of the levels of the formation of professional competencies in the experimental and control groups of mathematics trainee teachers has statistically significant differences due to the implementation of the developed method of using the cloud service CoCalc, which confirms the hypothesis of the study.

The expediency of introducing CoCalc cloud service to the process of teaching mathematics and informatics teachers is substantiated within the research, that accounted for certain peculiarities of content formation of several mathematical and informational disciplines, other approaches to solving classical problems. The CoCalc service was used to provide adaptive learning and personalized approach, organization of students' teamwork and it was used due to its free license. Differentiation of the tasks will help to include in the educational process more tools that are presented in the cloud environment: Chatroom, LaTeXDocument, ManageaCourse, TaskList, and not only the most common resource – the worksheet. The results of this application were approved by the pedagogical experiment, undertaken in Kryvyi Rih National University, Ukraine. It was proved that the level of students' professional competencies would be higher if to introduce into the learning process the proposed method of using CoCalc cloud service.

In 2019 the results of the different aspects of the study were tested at 28 scientific and practical events: 6 conferences (4 international ones); 17 workshops (1 international). The problematic issues of scientific research were discussed and presented for the scientific and pedagogical community by organizing and conducting by the authors a series of training sessions, seminars, webinars for scientific and pedagogical staff, and graduate students.

# 6 Conclusions and discussion

1.   It is advisable to include the components of the corporate and public clouds of the educational institution (databases and data collections, adaptive content management systems, cloud-based office software applications, specialized software training tools, language processing tools, educational robots, and others) as well as services of publicly available information systems (scientific-educational information networks and infrastructures, cloud-based educational, scientific services) into future teachers training.

2.   In the general model of the adaptive cloud-based system of education and professional development of teachers the components of the public and the corporate cloud of the educational institution are distinguished, as well as communication services and scientific and educational information networks. As now the hybrid cloud-based solutions are at demand there is no strict limitation of what kinds of the services would belong to any of the public or the corporate parts. The proposed model only provides the general structure and approach for the services supply. The important issue for further research is consider and build different configurations in view of the basic principles and approach. The experimental design was based on several available services for the different components of the general model that was outlined. Still the set of the services is not still exhaustive in any case.

3.   In the process of designing an adaptive cloud-oriented system of education and professional development of teachers of general secondary education, it is advisable to use the methodology of using the services of a scientific-educational cloud of an educational institution based on Microsoft Office 365, Google services and other; to use adaptive content management tools based on a public cloud; to implement methodology for supporting the adaptive knowledge-based processes of creation and use of electronic educational resources and other kinds of services within the outlined framework.

# References

1.   The EU explained: Digital agenda for Europe (2014). doi:10.2775/41229

2.   Teacher professional development (OECD, Education GPS, The world of education at your fingertips, 2019), https://gpseducation.oecd.org/revieweducationpolicies/#!node=41732&filter=all. Accessed 29 Feb 2020

3.   T.L. Mazurok, Yu. K. Todortsev, in *Proceedings of the First International Conference on the Adaptive Learning Management Technologies ATL-2015*, South Ukrainian National Pedagogical University named after K. D. Ushynsky, Odessa, 23–25 September 2015

4.   Z. Maamar et al., Int. J. Econom. Bus. Res. **5** (4), 1–21 (2009)

5.   V. Tataurov, M. Shyshkina, Sci. J. Phys. Math. Edu. 4 (22), 124–129 (2020)

6.   V. Bykov, Inf. Tech. Educ. **10**, 8–23 (2011)

7.   Yu. Nosenko, M. Popel, M. Shyshkina, CEUR Workshop Proceedings **2433**, 173–183 (2019)

8.   C. Goldberg, 10 Education Trends that will Shape the 2019-2020 Academic Year (Touro College, Online Education for Higher 2019), http://blogs.onlineeducation.touro.edu/10-education-trends-that-will-shape-the-2019-2020-academic-year/. Accessed 29 Feb 2020

9.   M. Bulger, Personalized Learning: The Conversations We're Not Having (Data&Society, 2016), https://datasociety.net/pubs/ecl/Personalized Learning_primer_2016.pdf. Accessed 29 Feb 2020

10.   C. Devendra, R. Eunhee, K. Jihie, in *Proceedings of the Third Annual ACM Conference on Learning at Scale*, University of Edinburgh, Edinburgh, 25–26 April 2016

11.   J.S. Groff, *Personalized Learning: The State of the Field & Future Directions* (Center for Curriculum Redesign, Boston, 2017)

12.   T.G. Mathewson, These 7 trends are shaping personalized learning (Education Dive, 2017), https://www.educationdive.com/news/these-7-trends-are-shaping-personalized-learning/434575/. Accessed 29 Feb 2020

13.   D. Newman, Top 6 Digital Transformation Trends In Education (Forbes, 2017), https://www.forbes.com/sites/danielnewman/2017/07/18/top-6-digital-transformation-trends-in-education/#af0e8082a9a2. Accessed 29 Feb 2020

14.   S. Vosniadou, *How Children Learn* (The International Academy of Education, Brussels, 2001), pp. 8–28

15.   Unlock Limitless Learning (Microsoft Education, 2020), https://www.microsoft.com/en-us/education/default.aspx. Accessed 29 Feb 2020

16.   J. Nakic, A. Granic, V. Glavinic, J. Edu. Comp. Res. **51**(4), 459–489 (2015)

17.   R. Shen, R. Richardson, in *Proceedings of Joint Conference on Digital Libraries 2003*, Rice University, Houston, 31 May 2003, ed. by A. Fox

18.   Knewton adaptive learning Building the world's most powerful recommendation engine for education (Knewton, 2015), https://cdn.tc-library.org/Edlab/Knewton-adaptive-learning-white-paper-1.pdf. Accessed 29 Feb 2020

19.   B. Upbin, Knewton is building the world's smartest tutor (Forbes, 2011), https://www.forbes.com/sites/bruceupbin/2012/02/22/knewton-is-building-the-worlds-smartest-tutor/. Accessed 29 Feb 2020

20. N. Sonwalkar, The First Adaptive MOOC: A Case Study on Pedagogy Framework and Scalable Cloud Architecture – Part I. MOOCs Forum (2013). doi:10.1089/mooc.2013.0007

21. Aspire Public Schools (Next Generation Learning Challenges, 2018), https://www.nextgenlearning.org/grantee/aspire-public-schools-1. Accessed 6 March 2020

22. D. Onah, J. Sinclair, in *Proceeding of the 9th International Technology, Education and Development Conference*, IATED, Madrid, 2–4 March 2015

23. S. Svetsky, O. Moravcik, SVK Utility Mod. UV 7340 (2016)